\documentclass[showpacs,prb,twocolumn]{revtex4}% Physical Review B

\usepackage{graphicx}% Include figure files
\usepackage{dcolumn}% Align table columns on decimal point
\usepackage{bm}% bold math

\begin{document}

\preprint {}

\title{Nanocrystallization and amorphization induced by reactive nitrogen sputtering in iron and permalloy}

\author{Rachana Gupta}

\author{Mukul Gupta}
\email{mukul.gupta@psi.ch}
\affiliation
{Laboratory for Neutron
Scattering, ETHZ $\&$ PSI,Paul Scherrer Institut,Villigen,CH-5232,
Switzerland}
\date{\today}% It is always \today, today,

\begin{abstract}
Thin films of iron and permalloy (Ni$_{80}$Fe$_{20}$) were
prepared using an Ar+N$_{2}$ mixture with magnetron sputtering
technique at ambient temperature. The nitrogen partial pressure,
during sputtering process was varied in the range of 0 $\leq$
R$_{N2}$ $\leq$ 100$\%$, keeping the total gas flow at constant.
At lower nitrogen pressures (R$_{N2}$ $\leq$ 33\%) both Fe and
NiFe, first form a nanocrystalline structure and an increase in
R$_{N2}$, results in formation of an amorphous structure. At
intermediate nitrogen partial pressures, nitrides of Fe and NiFe
were obtained while at even higher nitrogen partial pressures,
nitrides themselves became nanocrystalline or amorphous. The
surface, structural and magnetic properties of the deposited films
were studied using x-ray reflection and diffraction, transmission
electron microscopy, polarized neutron reflectivity and using a DC
extraction magnetometer. The growth behavior for amorphous film
was found different as compared with poly or nanocrystalline
films. The soft-magnetic properties of FeN were improved on
nanocrystallization while those of NiFeN were degraded. A
mechanism inducing nanocrystallization and amorphization in Fe and
NiFe due to reactive nitrogen sputtering is discussed in the
present article.
\end{abstract}

%PACS, the Physics and Astronomy
\maketitle

\section{\label{sec:level1}Introduction}
During recent years, nanostructured and amorphous thin films and
multilayers of magnetic materials have attracted tremendous
scientific and technological interests due to their unique
properties compared to conventional crystalline materials.\cite
{Angell_JAP00,Deben_Nature01,McHenry_PMS99,Dunlop_PRL03} In
nanocrystalline materials, as the grain size decreases, there is a
significant increase in the volume fraction of grain boundaries or
interfaces. This characteristic strongly influences the chemical
and physical properties of the material. In particular, a decrease
in the grain size results in improved soft-magnetic properties. On
the other hand, amorphous phases are expected to be free from
grains and grain-boundaries which often results in release of
intrinsic stresses, decrease in magnetic anisotropy and a smoother
surface or interface. Furthermore, grains or grain-boundaries act
like active path for diffusion, and therefore, atomic
self-diffusion in amorphous phases is expected to be lower.
Amorphous or nano grain thin films exhibit a short range ordering
in the microstructure and their structural, mechanical,
electrical, and magnetic properties often depends on the methods
and conditions of preparation.\cite
{Line_TSF96,Sulitanu_JOEAM03,Faupel_RMP03} Various attempts have
been made to achieve amorphization in binary or multicomponent
metal-metal and metal-metalloid systems using different techniques
such as rapid-melt quenching,\cite{Elliott_Amorph} mechanical
alloying,\cite{Koch_APL83} hydrogenation,\cite {Meng_APL88}
pressure,\cite{Sharma_PMS96} interdiffusion
reaction\cite{Schwarz_PRL83,Johnson_PMS85,Samwer_PR88} and ion or
electron irradiation.\cite{Holz_PRL83,Takeda_PRL99}

Quite recently, nitrogen reactive sputtering has also been used to
achieve a nanocrystalline or amorphous phase.\cite
{Hellgren_PR99,Babonneau_APL03,Gupta_PRB02,Gupta_JJPS04} In the
sputtering process the adatoms have energy of the order of few
tens of eV,and during condensation onto the substrate,adatoms are
quenched and may form an amorphous or fine grain structure. At the
same time, when sputtered using low Z reactive ions e.g. nitrogen
ions, they may occupy interstitial sites in the unit cell of
sputtered species, causing a distortion of the unit cell. A
combined effect of these situations may lead to a nanocrystalline
or amorphous structure of the deposited film. In order to verify
such a mechanism causing nanocrystallization or amorphization, two
different materials namely bcc Fe and fcc NiFe permalloy were
chosen for the present study. In earlier studies, Fe thin films
have been prepared using an Ar+N$_{2}$ gas mixture by magnetron
sputtering,\cite{Rissanen_JAC98,Russak_JAP91} rf
sputtering,\cite{Bobo_JAP95,Nie_TSF03}pulsed laser
deposition,\cite{Yoshitake_IEEETM95,Gupta_JAC01,Wang_JVSTA03}
ion-beam enhanced deposition (IBED),\cite{Guibin_SCT97} etc. The
motivation of most of these studies was to obtain nitrogen poor
Fe$_{16}$N$_{2}$ phase which possess very high magnetic
moment.\cite{Kim_APL72,Zhang_PRB96}In some of these studies an
amorphous or nanocrystalline phase of FeN was obtained at low
nitrogen pressure.\cite{Bobo_JAP95,Nie_TSF03,Guibin_SCT97}
However, a detailed investigation of evolution of nanocrystalline
or amorphous phases and a mechanism inducing nanocrystallization
or amorphization was not studied. It is known that when heated,
evaporated, ablated or sputtered in nitrogen environment or with
nitrogen ion, iron forms a microstructure with a variety of FeN
alloys and compounds, including the recently discovered
new-cubic-type nitrides.\cite{Rissanen_JAC98,Gupta_JAC01}
Ferromagnetic nitrides of iron have received tremendous interests
in magnetic functional devices.\cite{Chang_IEEETM87,Wang_JMMM04}On
the other hand, NiFe alloy with a composition of
Ni$_{80}$Fe$_{20}$ is a well-known soft-magnetic alloy and is
known as permalloy. It forms a face centered cubic structure of
the type Ni$_{3}$Fe. In a recent study by Chiba \emph{et
al.},\cite{Chiba_JMMM02} NiFe nitrides were deposited using rf
sputtering technique for a nitrogen flow in the range of 5-30\%. A
decrease in saturation magnetization is reported, however a
detailed variation in microstructure with higher nitrogen content
was not investigated.

In present work our aim is to explore the structural and magnetic
properties and growth behavior of the bcc Fe and fcc NiFe thin
films prepared using reactive nitrogen sputtering in the whole
nitrogen partial pressure range (0-100\%). In a study by Kawamura
\emph{et al.}\cite{Kawamura_Vac00} thin films of NiN were studied.
In the present case it was found that both Fe and NiFe, forms an
amorphous or nanocrystalline phase of either the element or a
nitride of them when sputtered with a nitrogen poor or rich
mixtures. Polycrystalline films containing a mixture of nitrides
were obtained at intermediate gas pressures and below or above,
the long range ordering of either the pure metal or nitrides of
them is restricted and a nanocrystalline or amorphous structure is
obtained. On the basis of obtained results a mechanism leading the
breakdown of long range ordering is discussed. In order to
understand the physical properties of the formed amorphous phases,
crystallization process was studied after annealing the thin films
in vacuum. It is known that amorphous films have a smoother
surface, due to absence of grains and lattice defects; the growth
behavior of an amorphous phase and for comparison of pure Fe and
nanocrystalline FeN was studied. Magnetic properties of
ferromagnetic films were studied using a DC extraction
magnetometer and in order to avoid diamagnetism of the substrate,
the magnetic moment was also determined using polarized neutron
reflectometry. The results of abovementioned studies are presented
and discussed in this article.

\section{\label{sec:level2}Experimental methods}
Thin films of Fe and permalloy (Ni$_{80}$Fe$_{20}$) were prepared
by magnetron sputtering using a gas mixture of Ar+N$_{2}$. The
nitrogen partial pressure, defined as, R$_{N2}$ =
P$_{N2}$/(P$_{N2}$ $+$ P$_{Ar}$)$\times$ 100\%, was varied at 0,
2, 5, 10, 20, 33, 50, 83 and 100\% for Fe and 0, 5, 10, 20, 33,
50, 59, 83 and 100\% for NiFe. The gas flows in the vacuum chamber
were controlled using mass flow controllers and the total gas flow
for sputtering was kept fixed at 10 cm$^{3}$/min. Circular targets
of pure Fe or permalloy, 75 mm in diameter, were sputtered with
the gas mixture. A constant sputtering power of 50 W was used in
all depositions. The targets were covered with slits of width 80mm
to restrict the plasma. The cathode (target) and the substrate
were mounted parallel to each other at a distance of about 8 cm.
Before depositions a base vacuum of the order of
1$\times10$$^{-6}$ mbar was obtained and the vacuum chamber was
flushed with Ar and N$_{2}$ gas so as to avoid contamination of
other gases inside the vacuum chamber. The pressure during
deposition was in the range of 4-8$\times10$$^{-3}$ mbar. The
substrates were mounted below the targets and oscillated with
respect to the central position of the target for better
uniformity of the deposited samples. All the samples were
deposited at room temperature ($\sim$298K, without intentional
heating) on float glass or Si (100) substrates. Thin films for
growth studies were deposited in a single sputtering run (for one
composition) onto a glass substrate. The targets were covered with
a small slit of size 15 mm and the substrate was exposed to the
center of target for different amount of time to obtain different
thicknesses. The substrate was translated using a computer
controlled linear translation stage. The thicknesses of the films
were determined using x-ray reflectivity (XRR) technique and the
structure of the films was investigated using grazing-incidence
x-ray diffraction (XRD) using Cu K$\alpha$ x-rays. For all the
measurements the incident angle was kept fixed just above the
critical angle of the film, to minimize the background due to
diffraction of the substrate. The bulk magnetic measurements were
performed using a DC extraction magnetometer with the magnetic
field applied parallel to the plane of the film using a physical
property measurement system (PPMS). In order to determine the
magnetic moment of the films, independent of substrate magnetism
or sample area, the polarized neutron reflectivity (PNR)
measurements were performed at the saturation field of the
samples. The measurements were performed at fixed angle of
incidence in the time of flight (ToF) mode at
AMOR(SINQ/PSI).\cite{Gupta_PramJP04}

\section{\label{sec:level3}Results and Discussion}
\subsection{\label{sec:leve31}Structural properties : FeN}
\begin{figure}
\includegraphics[width=80mm,height=100mm]{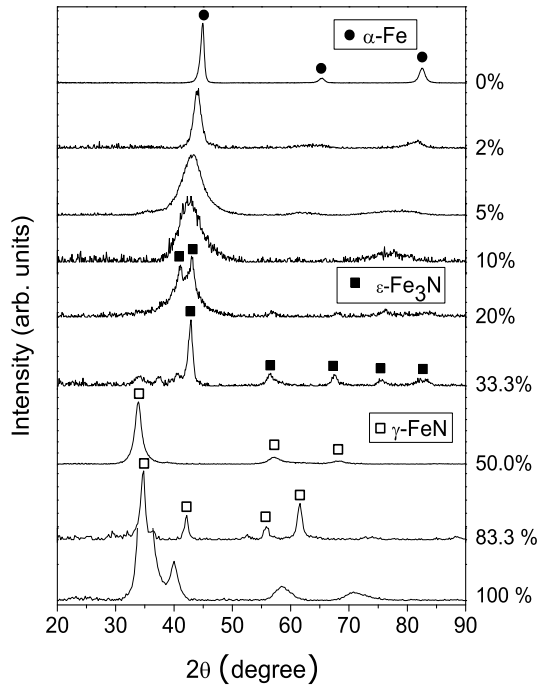}
\caption{\label{fig:Fig1} Grazing incidence x-ray diffraction
pattern of FeN thin films prepared with different nitrogen partial
pressure.}
\end{figure}

Fig.~\ref{fig:Fig1} shows the grazing-incidence XRD pattern of Fe
films prepared with different nitrogen partial pressure. The film
prepared with Ar gas only, shows reflections corresponding to bcc
$\alpha$-Fe with orientation in the direction of (110) plane. For
the films prepared with 2\%-20\% nitrogen partial pressure (rest
Ar), the XRD pattern shows a structure similar to bcc $\alpha$-Fe,
however all the peaks were broad and the peak positions were
shifted to lower angle side as compared to the XRD pattern of pure
Fe film. The line width of the diffracted pattern can be used to
calculate the grain size of the diffracting specimen in the
direction perpendicular to the plane of the film using Scherrer
formula,\cite{Cullity_XRD} $t$ = 0.9$\lambda$/$b$ $\cos$ $\theta$,
where $t$ is the grain size, $b$ is an angular width in terms of
2$\theta$, $\theta$ is the Bragg angle and $\lambda$ is the
wavelength of the radiation used. For the film prepared with Ar
gas only, the average grain size was 13$\pm$1 nm while in the
presence of 2\% nitrogen partial pressure during sputtering, it
decreases to 6$\pm$1 nm; half of the value found without any
nitrogen. This result indicates that even the presence of nitrogen
as small as 2\%, significantly affects the growth of Fe crystals
and restricts the long-range ordering. The positions of Bragg
peaks for the sample prepared with 2\% nitrogen, were shifted to
lower-angle side, indicating an increase in the inter-atomic
spacing. The average interatomic distance can be estimated using
the relation $a$ = 1.23/2 $ \sin$$\theta$, where $\theta$ is taken
to be the angle at the center of the peak, and the factor 1.23 is
a geometric factor which rationalizes the nearest neighbor
distance with the spacing between ``pseudo-close packed
planes".\cite{Guinier_XRD} Comparing the interatomic spacing for
the film prepared without nitrogen and with 2\% nitrogen, the
average interatomic spacing was found to be expanded by 2\%.
On increasing the nitrogen partial pressure the width of the
reflections further increases and peak position continues to shift
towards the lower angle side. At 5 and 10\% nitrogen partial
pressure the line width of the peak becomes as large as
4$^{\circ}$, which is close to the value found for conventional
iron based amorphous alloys.\cite{McHenry_PMS99}

The amorphous nature of the film deposited at R$_{N2}$ =10\%, was
confirmed with transmission electron microscopy (TEM). A thin film
of thickness 70 nm was directly deposited on a carbon coated TEM
grid.
\begin{figure}[!ht]
\includegraphics[width=95mm,height=80mm]{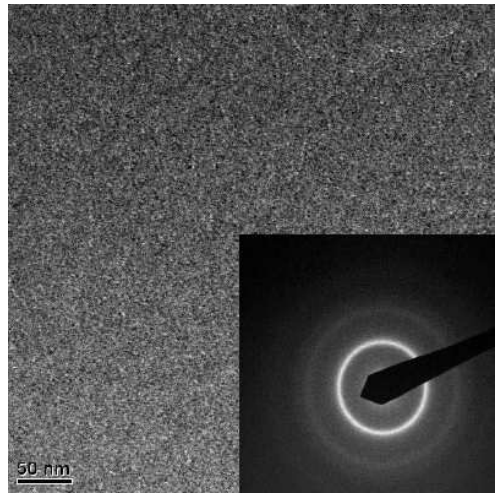}
\caption{\label{fig:Fig2} TEM planar view of the sample prepared
at 10\% nitrogen partial pressure. The inset of the picture shows
the electron diffraction pattern.}
\end{figure}
Fig.~\ref{fig:Fig2} shows a representative TEM micrograph along
with the electron diffraction pattern. Similar micrographs were
observed through out the plane of the film. The micrograph
essentially showed a feature less structure and the electron
diffraction pattern showed diffuse diffraction ring which confirms
the amorphous nature of the film.

\begin{figure}
\includegraphics [width=75mm,height=65mm]{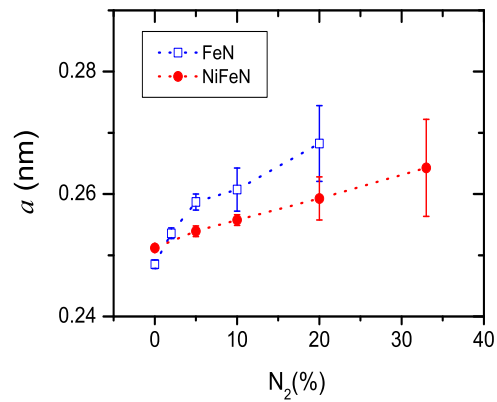}
\caption{\label{fig:Fig3} Inter-atomic spacing as a function of
nitrogen partial pressure in FeN and NiFeN.}
\end{figure}

Fig.~\ref{fig:Fig3} shows a plot of average interatomic distance
$a$, as function of increase in the nitrogen partial pressure in
the range of 0-20\% for FeN, and 0-33\% for NiFeN. As can be seen
from the figure, with an increase in the amount of nitrogen, the
interatomic spacing continues to increase. However at 20\% the
broad hump overlaps with two sharp peaks. The peak positions of
the sharp peaks correspond to hcp $\epsilon$-Fe$_{3}$N phase. And
the overall structure can be considered as a mixture of amorphous
bcc-Fe along with hcp $\epsilon$-Fe$_{3}$N phase. At 33.3\%
nitrogen partial pressure the structure changes completely and
$\epsilon$-Fe$_{3}$N phase along with $\zeta$-Fe$_{2}$N phases
were obtained. On further increasing the nitrogen partial pressure
at 50\%, the structure changes again reflections corresponding to
new-cubic-type phase were obtained. It may be noted that width of
the Bragg peak at 34$^{\circ}$ is
(1.4$^{\circ}$$\pm$0.01$^{\circ}$) corresponding to an average
grain size of about 6 nm, which is an indication of formation of a
nano grain structure. At 83.3\% nitrogen partial pressure, sharp
peaks corresponding to $\gamma$$^{\prime\prime\prime}$-FeN phase
were observed. On further increasing the nitrogen partial pressure
to 100\%, the peak widths again starts increasing, indicating
re-formation a nanocrystalline structure.

\subsection{\label{sec:level32}Structural properties : NiFeN}
The permalloy target was also sputtered with a mixture of
Ar+N$_{2}$ by varying the nitrogen partial pressure in the range
of 0-100\%.
\begin{figure}[!hb]
\includegraphics [width=85mm,height=130mm]{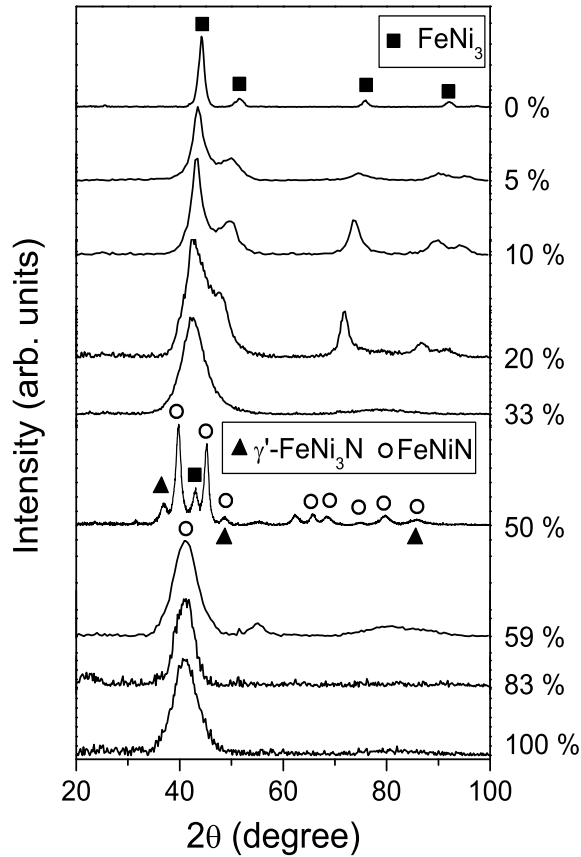}
\caption{\label{fig:Fig4}Grazing incidence x-ray diffraction
pattern of NiFeN thin films prepared with different nitrogen
partial pressure.}
\end{figure}
Fig.~\ref{fig:Fig4} shows grazing incidence x-ray diffraction
pattern of NiFeN thin films prepared at different nitrogen partial
pressure. The film prepared with Ar gas only shows reflections
corresponding permalloy phase as indexed in the figure. As the
nitrogen partial pressure is increased, the reflection starts
broadening and the reflection with indices (111) and (200) starts
merging together. A clear shift in the positions of Bragg peak is
also evident. The grain size for the film sputtered with Ar only
was 7 nm which decreases to 3.5 nm after sputtering with 5 or 10\%
nitrogen. With a further increase in the nitrogen partial
pressure, an amorphous phase appeared at 33\%. As compared to FeN,
the shifts in the positions of Bragg peaks were rather small (see
fig.~\ref{fig:Fig3}). Also overall increase in the interatomic
distance at similar nitrogen pressure was smaller in NiFe as
compared to Fe. A discussion related to this issue is given in
section 3. While comparing the observed results with that of NiN
studied by Kawamura \emph{et al.},\cite{Kawamura_Vac00} similar
broadening and expansion of unit cell of Ni was observed. On
increasing the R$_{N2}$ to 50\% with NiFe, the structure was
changed completely and several peaks were observed in the XRD
pattern. The phase identified at this pressure is a mixture of
FeNiN + $\gamma$$^{\prime}$-FeNi$_{3}$N. On further increasing
R$_{N2}$ to 59\% or above, a broad hump around 2$\theta$ =
40$^{\circ}$ appears along with faint reflections at higher
angles. This hump appears to be an envelop of several reflection
observed for R$_{N2}$ = 50\% sample and indicates re-amorphization
of the polycrystalline permalloy nitride structures formed at
R$_{N2}$ = 50\%.

Nanocrystallization or amorphization induced by reactive nitrogen
sputtering in Fe, Ni and NiFe can be explained with a single
mechanism. At low nitrogen partial pressures, nitrogen ions does
not react with Fe, Ni or NiFe and nitrogen is incorporated in the
interstitial sites, making an expansion of the unit cell. At
intermediate nitrogen pressure, a chemical reaction between
nitrogen and Fe, Ni or NiFe is favorable which results in
formation of nitride phases. At further higher nitrogen pressures,
deformation of the formed nitride phase starts and the end
structure is again nanocrystalline or amorphous. Detailed
mechanism inducing nanocrystallization or amorphization is
discussed in section 3.

\subsection{\label{sec:level33}Crystallization behavior of amorphous films}

From the observed XRD results it is evident that when nitrogen
partial pressure during sputtering is 5 and 10\%, an amorphous
phase of FeN was formed while at 33\% and above 50\%, amorphous
NiFeN phase was formed. It would be interesting to study the
crystallization behavior of these amorphous phases in order to
understand their properties. Two sets of samples were chosen for
crystallization studies (i) Fe-rich, FeN samples prepared at
R$_{N2}$ = 2, 5, 10 and 20\% and (ii) N-rich, NiFeN sample
prepared at R$_{N2}$ = 83\%. These films were annealed in a vacuum
furnace isochronally for 1 hour at each temperature. In order to
avoid the fluctuations in temperature all the four FeN films were
annealed simultaneously in the vacuum furnace.
\begin{figure*}[!ht]
\includegraphics[width=70mm,height=60mm]{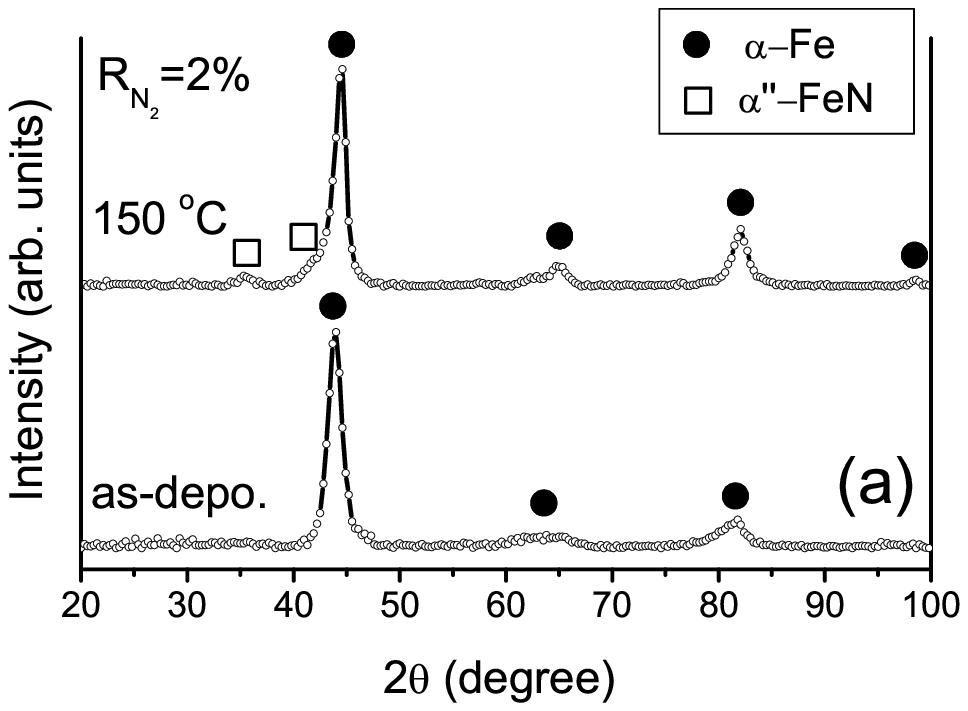}
\includegraphics[width=70mm,height=60mm]{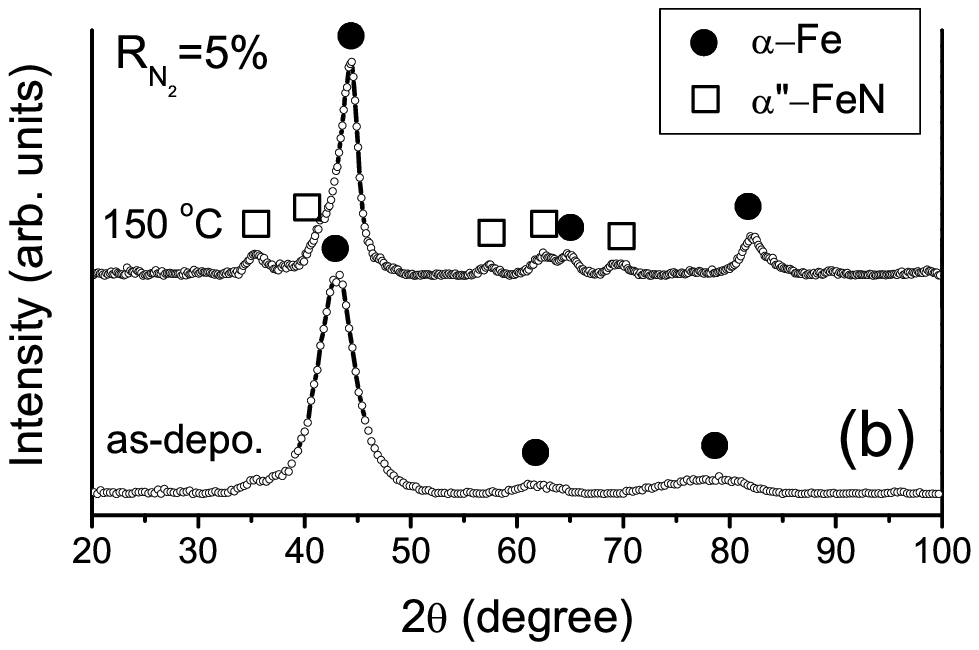}
\includegraphics[width=70mm,height=60mm]{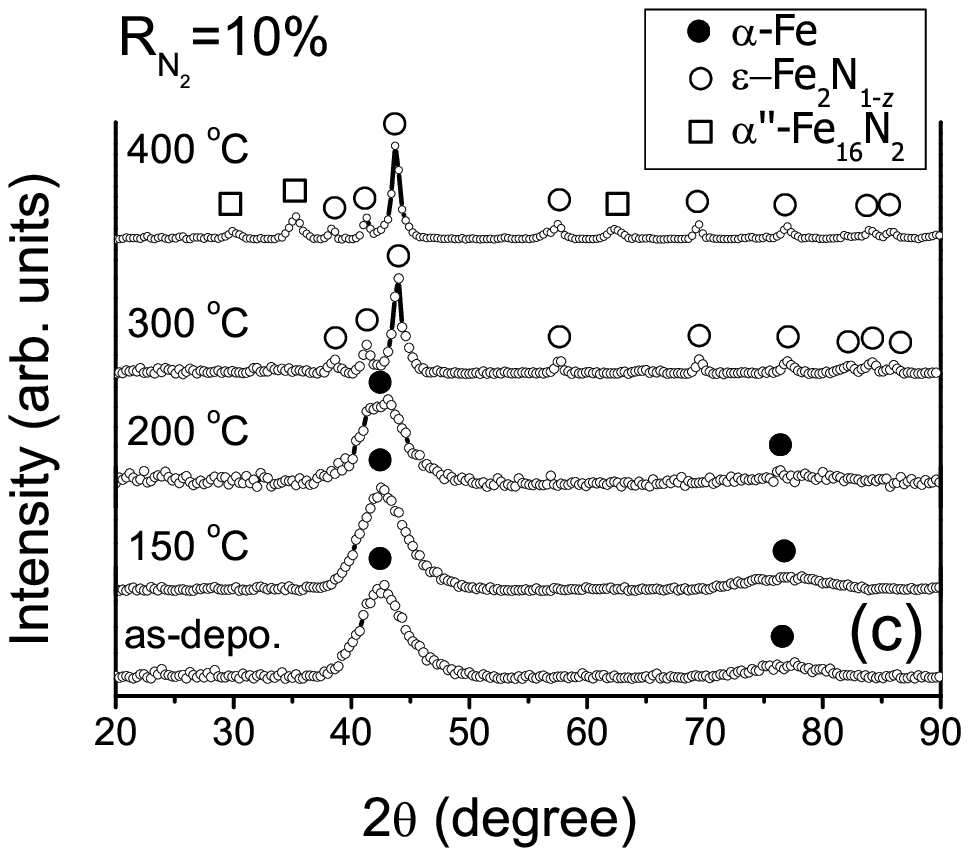}
\includegraphics[width=70mm,height=60mm]{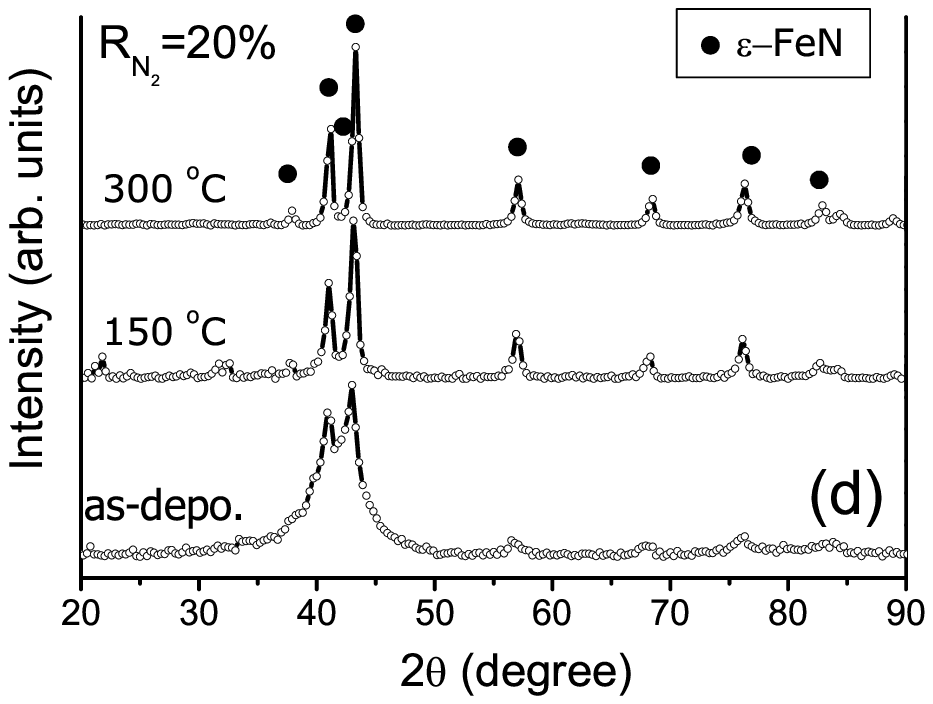}
\caption{\label{fig:Fig5} Grazing incidence x-ray diffraction
pattern of FeN film prepared with 2\% (a), 5\% (b), 10\% (c) and
20\% (d), in the as-deposited state and after vacuum annealing at
various temperatures. The films were annealed isochronally for 1
hour. }
\end{figure*}
Fig.~\ref{fig:Fig4}(a-d) shows the grazing incidence XRD pattern
of FeN annealed films. It is interesting to observe that the films
prepared with 2 and 5\% nitrogen partial pressure were highly
unstable, and even at 150 $^{\circ}$C, Bragg peaks corresponding
to $\alpha$$^{\prime\prime}$-FeN appear in the XRD pattern. The
film prepared with R$_{N2}$ = 10\%, was found to be more stable
and it remained amorphous up to an annealing temperature of 200
$^{\circ}$C. On further annealing at 300 $^{\circ}$C the amorphous
hump splits into three sharp peaks corresponding to
$\epsilon$-Fe$_{2}$N$_{1-z}$ phase and at 400 $^{\circ}$C, new
peaks corresponding to $\alpha$$^{\prime\prime}$-Fe$_{16}$N$_{2}$
phase were observed. For the film prepared with R$_{N2}$ = 20\%,
the XRD pattern reveals a composite structure consisting of
amorphous bcc-Fe and $\epsilon$-Fe$_{2}$N$_{1-z}$ phase in the
as-deposited state and after annealing at 150 $^{\circ}$C, the
structure crystallizes into $\epsilon$-Fe$_{2}$N$_{1-z}$ phase.
Observed crystallization behavior of amorphous phase shows that
the amorphous phase shows a better stability for the film prepared
with R$_{N2}$ = 10\%, and the films prepared either with higher or
lower nitrogen are formed in a highly metastable state.
\begin{figure}[ht]
\includegraphics[width=70mm,height=70mm]{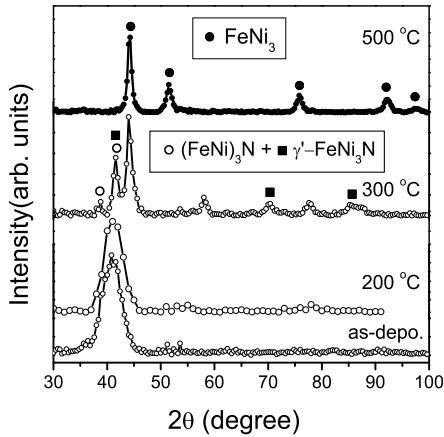}
\caption{\label{fig:Fig6} Grazing incidence x-ray diffraction
pattern of NiFeN thin film prepared with 83\% nitrogen partial
pressure after annealing at different temperatures. }
\end{figure}
Fig.~\ref{fig:Fig6} shows grazing incidence XRD pattern of the
NiFeN film prepared at R$_{N2}$ = 83\%, after annealing at
different temperatures. As can be seen from the figure, up to an
annealing temperature of 250 $^{\circ}$C, no significant changes
in the XRD pattern were observed. While after annealing at 350
$^{\circ}$C, several peaks were observed. The most intense peaks
corresponds to (FeNi)$_{3}$N and $\gamma$$^{\prime}$-FeNi$_{3}$N
phases. Smaller peaks at 53.6, 58.2 and 77.8$^{\circ}$ could not
be identified. After further annealing at 500 $^{\circ}$C, no
nitride phase was observed, indicating out diffusion of nitrogen.

The observed crystallization behavior of both amorphous FeN and
NiFeN phases is different as compared to iron based binary or
multi-component alloys. In conventional metal-metal amorphous
alloys, generally crystallization occurs in 2 steps, in the first
step a nanocrystalline microstructure co-exists with parent
amorphous phase whereas in the second step an intermetallic
compound along with nanocrystalline phase precipitates out. The
nominal reaction for such crystallization process had been given
as: amorphous $\rightarrow$ $\alpha$+amorphous $\rightarrow$
$\alpha$+ $\beta$; where $\alpha$ is the primary phase that
precipitates out from the amorphous matrix and $\beta$ is an
intermetallic compound.\cite{Hono_MC00,Zhu_JPDAP04} In the present
case, however, crystallization takes place in a single step and
annealing at temperatures above crystallization temperatures
essentially results in nitrogen out diffusion. The amorphous
structure remained amorphous up to certain annealing temperature
and thereafter mixed nitride phase were observed. On further
annealing due to nitrogen out diffusion pure metallic or nitrogen
poor phases are obtained.

\subsection{\label{sec:level34}Surface properties and growth behavior of FeN films}

The thickness of FeN and NiFeN thin films deposited for R$_{N2}$ =
0-100\%, were determined using x-ray reflectivity technique. Since
both FeN and NiFeN thin films were deposited in a similar manner,
detailed surface and growth behavior of FeN thin films was
investigated.
\begin{figure}[ht]
\includegraphics[width=85mm,height=100mm]{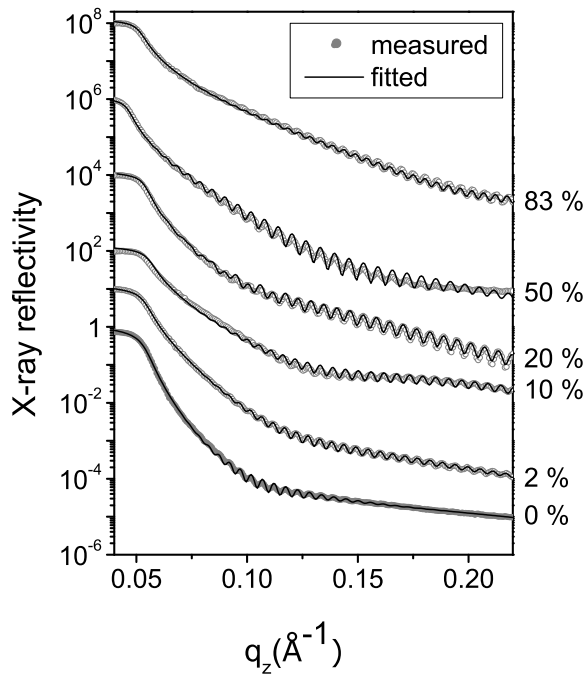}
\caption{\label{fig:Fig7} X-ray reflectivity pattern of FeN thin
films prepared with different nitrogen partial pressure.}
\end{figure}
Fig.~\ref{fig:Fig7} shows x-ray reflectivity pattern of FeN thin
films prepared at different nitrogen partial pressures. The x-ray
reflectivity pattern was fitted using a computer
program\cite{Parratt32} based on Parratt's
formalism.\cite{Parratt_PR54} Oscillations due to total thickness
of the films can be clearly seen in the reflectivity pattern.
\begin{table}
\caption{\label{tab:table1} Fitted x-ray reflectivity parameters
for FeN thin films prepared at different nitrogen partial
pressures.}
\begin{ruledtabular}
\begin{tabular}{cccccccc}
$R_{N2}$&$film$&$film$&$sub$ \\
  &thickness&roughness&roughness \\
(\%)& nm($\pm$0.2)& nm($\pm$0.1)& nm($\pm$0.2)\\
\hline 0& 103.5 & 3.3 & 0.6 \\
2& 105.4 & 2.0 & 0.6 \\
10& 101.6 & 0.4 & 0.6 \\
20& 98.3 & 1.0 & 0.6 \\
50& 93.5 & 1.0 & 0.6 \\
83& 113.5 & 0.8 & 0.6 \\
\end{tabular}
\end{ruledtabular}
\end{table}
The thickness of the films obtained after fitting the pattern and
was found in the range of 90-100 nm. The fitted parameters are
given in table~\ref{tab:table1}. A detailed fitting of the pattern
revealed that a thin layer with density about 50\% of the bulk of
the layers is formed on the surface. Such a layer may be formed
due to ``oxidation" of the surface when exposed to atmosphere. The
thickness of this layer was typically 2-3 nm. It is interesting to
see that the roughness of the film prepared with Ar gas only, was
3.3 nm, which decreases to 2 nm at R$_{N2}$ = 2\%, and was only
0.4 nm for R$_{N2}$ = 10\%. At R$_{N2}$ = 20 and 50\% the
roughness again increases to 1 nm and at 83\% again it decreases
slightly. While looking at microstructure of the deposited film
obtained from XRD measurements, there is a clear indication that
an amorphous phase is formed at lower and higher nitrogen partial
pressures. A decrease in the roughness of film is not unexpected
since amorphous structure is free of grains and lattice defects,
which may result in formation of a smoother surface or interface.
With the observed decrease in the roughness of the film for
amorphous samples, it would be interesting to study the growth
behavior of these films. For this purpose a series of FeN thin
films were deposited at R$_{N2}$ = 0, 10 and 83\%. In order to
minimize the parameters influencing the growth of a thin film, all
the film with one composition were prepared in a single sputtering
run. For this purpose, the area of sputtering target was masked
with a small slit of size 15 mm and all the films were prepared on
a glass substrate by exposing the substrate for different times to
the plasma at different positions on the substrate to obtain
different thicknesses.
\begin{figure*}
\includegraphics[width=80mm,height=100mm]{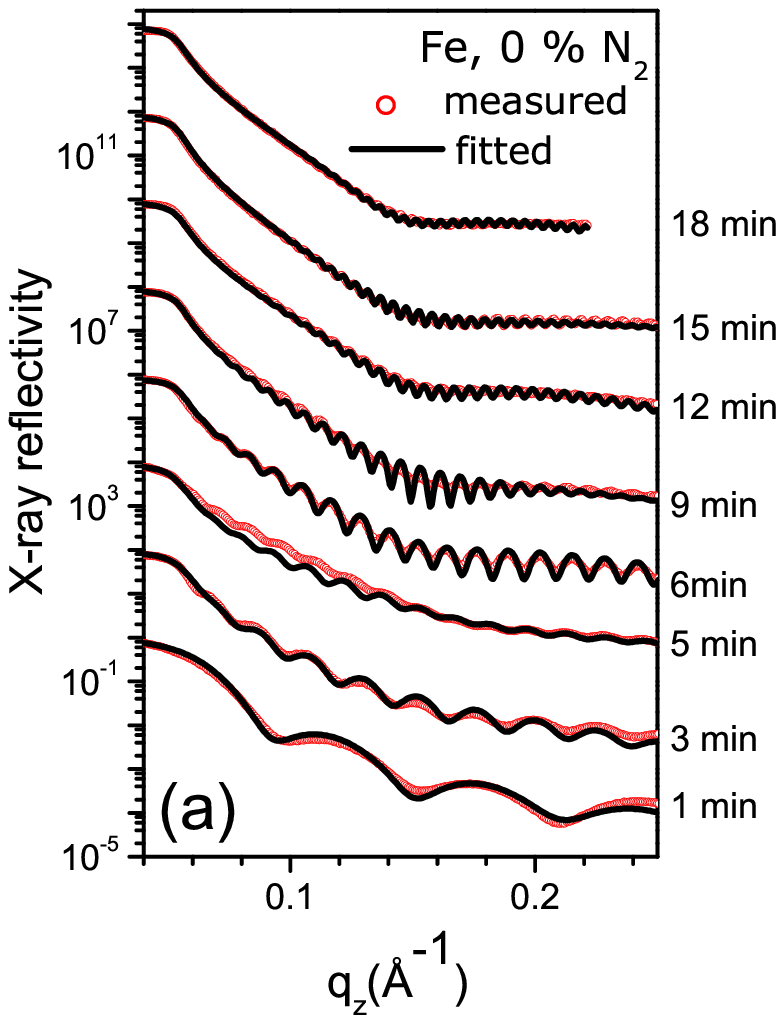}
\includegraphics[width=80mm,height=100mm]{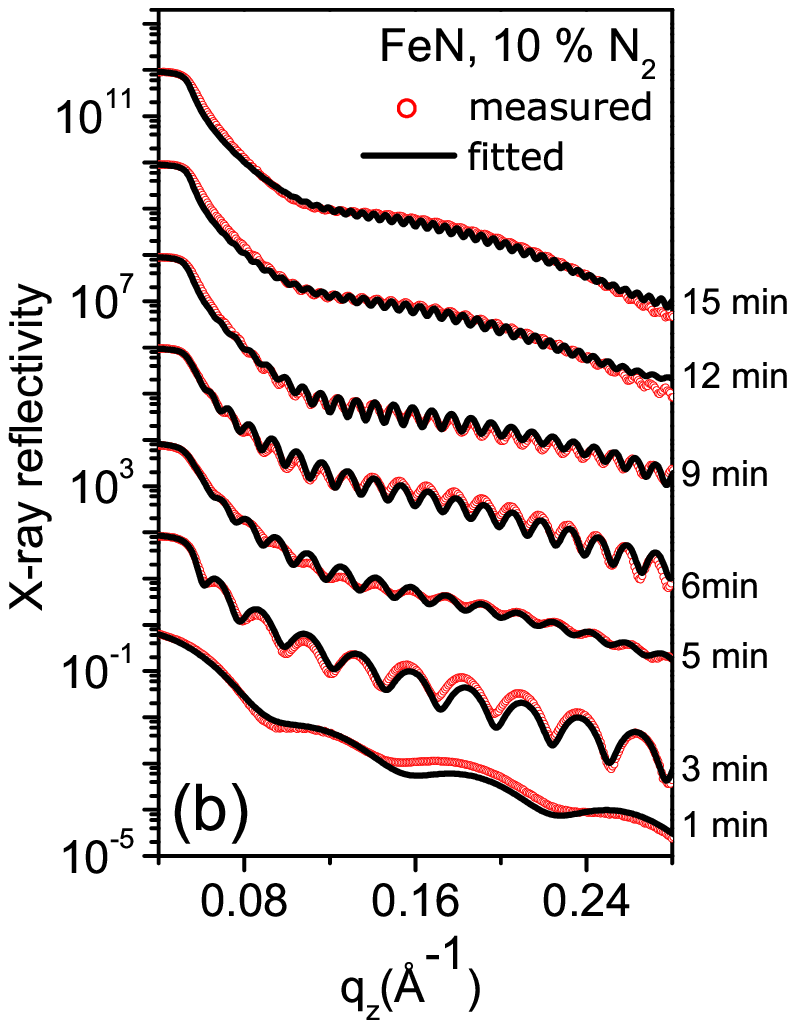}
\includegraphics[width=80mm,height=100mm]{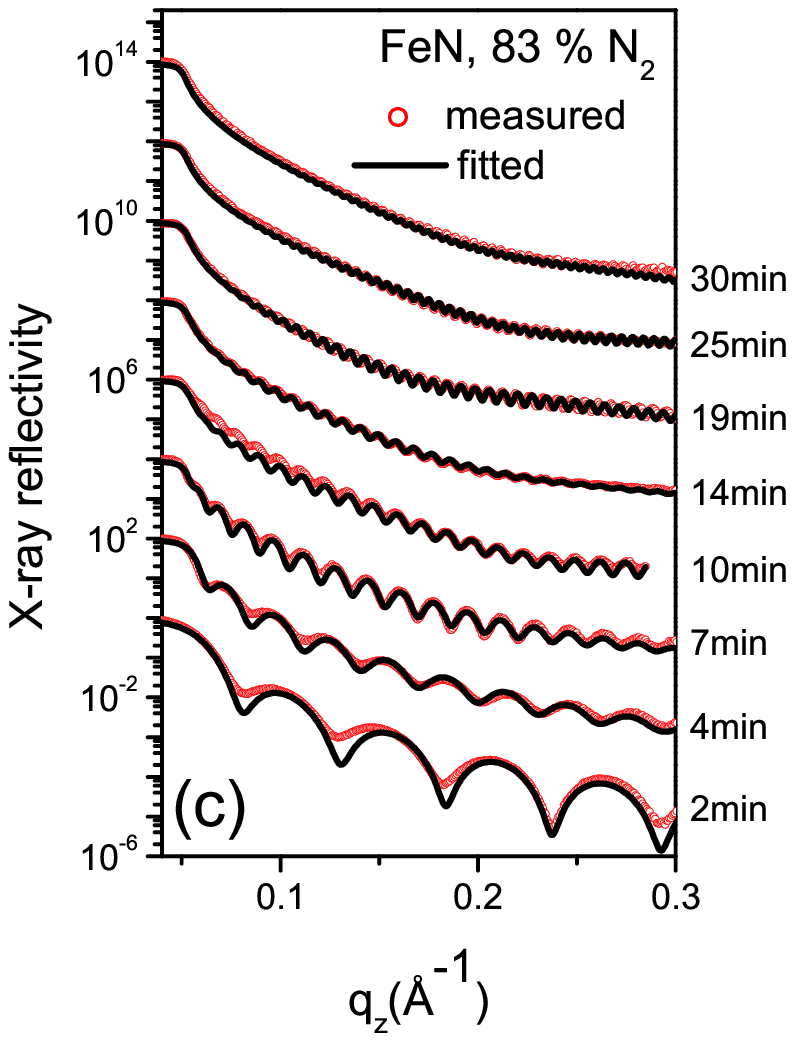}
\includegraphics[width=80mm,height=65mm]{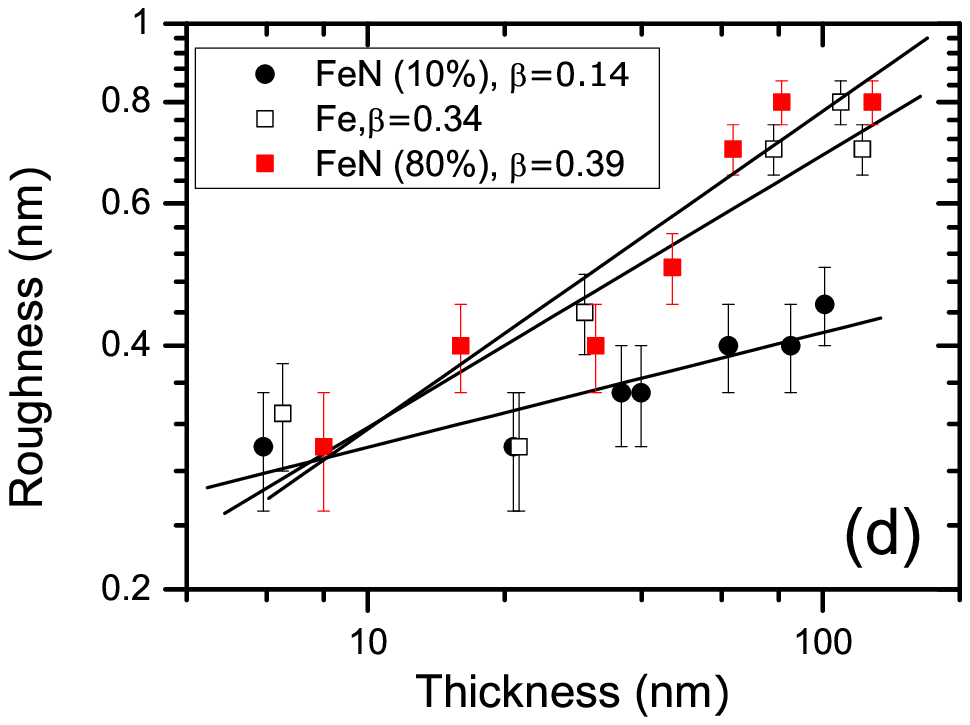}
\caption{\label{fig:Fig8} X-ray reflectivity pattern of pure Fe
film (a), FeN film with 10\% N (b) and FeN film with 83\% nitrogen
partial pressure (c). The evolution of film roughness with film
thickness (d).}
\end{figure*}

All the films were prepared in the thickness range of 10-150 nm.
X-ray reflectivity measurements on all the films were performed in
specular and off-specular mode. The off-set in off-specular
measurements was taken at the onset of specular reflection peak in
the rocking scan. This offset was 0.05$^{\circ}$. The off-specular
data was subtracted from the specular data to obtain the
``true-specular" data. Fig.~\ref{fig:Fig8}(a-c) shows the XRR
patterns of abovementioned thin films. The patterns were fitted
using a procedure as describe earlier. It was observed that for
pure iron film, the surface roughness increases monotonically with
an increase in the thickness while for the amorphous FeN film the
roughness of the film increases at a very slow rate. For the case
when FeN forms a nanocrystalline nitride at R$_{N2}$ = 83\%, the
surface roughness again show an increase with the thickness.
Previous studies have shown that the rms roughness ($\sigma$)
exhibits a power-law behavior,\cite{Gupta_JJPS04,Barabasi_grow} as
a function of the film thickness, given as: $\sigma$ $\sim$
t$^{\beta}$, accordingly, a double logarithmic plot of the rms
roughness of the films versus the film thickness should yield a
linear relation. Fig.~\ref{fig:Fig8}(d) shows a plot rms roughness
versus film thickness for the abovementioned samples and a
straight line fit to the data yields the roughness growth
exponent,$\beta$. For pure Fe film $\beta$ = 0.35, for amorphous
FeN film prepared at R$_{N2}$ = 10\%, $\beta$ = 0.14, and for
nanocrystalline FeN film prepared at R$_{N2}$ = 83\%, $\beta$ =
0.39. The value of $\beta$ for amorphous FeN is very close to that
obtained for amorphous SiO$_{2}$\cite{Lutt_PRB97} and nitrogen
rich amorphous FeN prepared using ion-beam
sputtering.\cite{Gupta_JJPS04} Under the various growth models
described in literature, it may be observed that for the values of
the roughness growth exponent lying in the range of 0.1-0.25, the
growth can be well described by the KPZ model, first introduced by
Kardar, Parisi, and Zhang.\cite{KPZ_PRL86} This type of the growth
processes takes into account a random deposition and a limited
relaxation of the particles at the surface. For the present case
the value of for amorphous sample lies well in this range, while
for both pure Fe and nanocrystalline FeN, the value of $\beta$ is
higher. A higher value of $\beta$ indicates non KPZ type growth
which is often observed for polycrystalline elements. The obtained
results clearly show a different growth mechanism for amorphous
film and support the argument that amorphous films yield smoother
surfaces. It may be noted that the absolute roughnesses of the
films prepared with smaller slits (15 mm) in front of the targets
were smaller as compared with the larger slits (80 mm).

\subsection{\label{sec:level35}Bulk magnetization measurements of FeN and NiFeN films}

\begin{figure}[!ht]
\includegraphics[width=85mm,height=65mm]{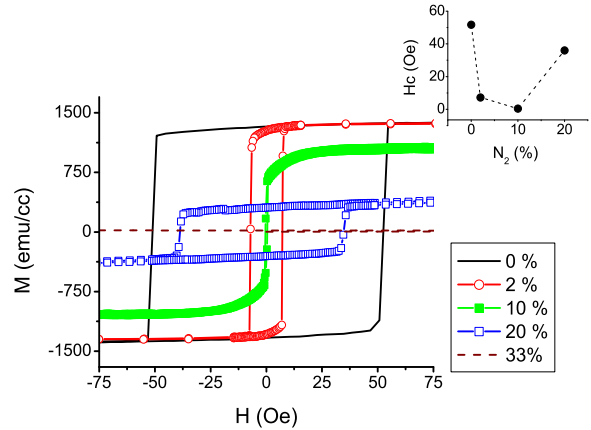}
\caption{\label{fig:Fig9}Magnetization measurements of FeN thin
films prepared at different nitrogen partial pressures. The inset
shows a variation in coercivity as a function of nitrogen partial
pressure during sputtering.}
\end{figure}

Fig.~\ref{fig:Fig9} shows MH curve of the FeN samples prepared at
different nitrogen partial pressure. For the film prepared with Ar
gas only, the behavior is as-expected for iron, and the value of
saturation magnetization (Ms) was almost equal to reported values
for bulk Fe. For the sample prepared at R$_{N2}$ = 2\%, the MH
loop shape changes to a typical soft-magnetic. While the value of
saturation magnetization is almost equal to pure Fe, the MH curve
shows a significant decrease in the value of coercivity (Hc).
It may be noted that the average grain size (D) for pure Fe is
about 13 nm, while for the sample prepared with 2\% nitrogen
partial pressure, the grain size reduce to 6 nm. A decrease in Hc
in nanocrystalline ferromagnets is expected as envisaged in the
random-anisotropy model
(RAM).\cite{Herzer_IEEE89,Herzer_IEEE90,Loffler_PRB98} It may be
noted that the ferromagnetic exchange length (L$_{ex}$) for
$\alpha$-Fe is 15-23 nm. \cite{Loffler_PRL2000,Gao_JAP2003}For
grain sizes D $<$ L$_{ex}$, Hc decreases with a decrease in grain
size by D$^{6}$. A reduction in grain size below ferromagnetic
exchange length allows exchange coupling between the neighboring
grains and results in a reduced effective anisotropy $\langle K
\rangle$. In the present case, however, the decrease in Hc was not
found to vary with D$^{6}$, rather it follows D$^{2-3}$ type
behavior. Since in the present case thickness of the films is
small, it is likely that effective averaging would be only in the
plane (area) of the film. In this situation the ferromagnetic
correlation volume would be proportional to L$_{ex}$$^{2}$ only,
in contrast to L$_{ex}$$^{3}$, in case of bulk materials. This
would reduce the number of grains over which the averaging is
done, and therefore a reduction in the magnetic anisotropy and Hc
is not expected to vary as D$^{6}$ as observed in nanocrystalline
ribbons or powders. Assuming averaging over N$ =
$(L$_{ex}$/D)$^{2}$ number of grains, the effective anisotropy
would be:
\begin{equation}
\langle K \rangle = { \frac{{K_{1}^{2}}{D^{2}}}{A} },
\end{equation}
where K$_{1}$ is magneto-crystalline anisotropy of the grains and
A is the exchange stiffness. This would mean that in case of a
thin film, Hc would follow D$^{2}$ type behavior rather than
D$^{6}$, as pointed out by Hoffmann \emph{et
al.}\cite{Hoffmann_JMMM93} The observed decrease in Hc, for the
sample prepared in presence of 2\% nitrogen partial pressure can
be understood accordingly.

On the other hand when nitrogen partial pressure was increased to
10\%, the alloy forms an amorphous structure and the magnetic
measurements show a decrease in Ms as well as Hc. A decrease in Hc
can be understood within RAM, when averaging is done on very fine
grains and magnetization follow the easy direction of each
individual grain. The decrease in the value of Ms can be
understood due to weakening of the exchange coupling between the
grains. Further, at R$_{N2}$ = 20\%, the Hc increase abruptly
while Ms continues to decrease.

It may be noted that at this nitrogen partial pressure amorphous
phase co-exists with hcp-$\epsilon$-Fe$_{3}$N phase. In case
$\epsilon$-Fe$_{3}$N phase is nonmagnetic (as found at higher
R$_{N2}$), presence of a nonmagnetic phase among the ferromagnetic
fine grains would result in a decrease in exchange length, which
in accordance with the RAM, causes anisotropy and coercivity to
increase because of incomplete averaging-out of random
anisotropies of grains within the exchange volume. The observed
increase in the Hc and decrease in Ms can be understood with this
argument. At further higher nitrogen partial pressure where
$\epsilon$-Fe$_{3}$N phase along with $\zeta$-Fe$_{2}$N phases
were obtained, the magnetization was almost zero, indicating
nonmagnetic nature of the film at this nitrogen pressure. The
films deposited at even higher R$_{N2}$\ were also nonmagnetic.

Crystallization behavior of the amorphous film deposited at
R$_{N2}$ = 10\%, was also studied with magnetization measurements
of the samples annealed at different temperatures.
\begin{figure}[!ht]
\includegraphics[width=85mm,height=70mm]{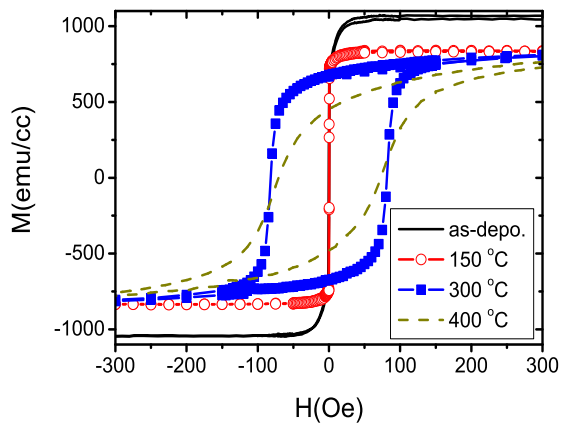}
\caption{\label{fig:Fig10}Magnetization measurements of FeN thin
film prepared at 10\% nitrogen partial pressure as function of
annealing temperature. }
\end{figure}
Fig.~\ref{fig:Fig10} shows magnetization measurements after
annealing at different temperatures as discussed in the previous
section. The M-H loop shape up, before crystallization temperature
was similar to that in the as-deposited state. The only
appreciable change was the shape of the M-H loop, which became
more square after annealing. Such a change in the M-H loop shape
is directly related to removal of strains which might have
developed during the deposition. At higher annealing temperature
where crystallization of amorphous phase takes place, the loop
shape was completely different. There was a sharp increase in the
value of coercivity and the average value of Ms decreased. At
further higher temperature, the value of Ms further decreased and
the loop shape looks broader. From the XRD measurements it is
evident that upon crystallization $\epsilon$-Fe$_{3}$N phase
precipitates out and an increase in the Hc and decrease in Ms is
similar as observed for the sample prepared at R$_{N2}$ = 20\%,
where $\epsilon$-Fe$_{3}$N phase existed. The observed
magnetization behavior can be understood accordingly.

\begin{figure}
\includegraphics[width=85mm,height=70mm]{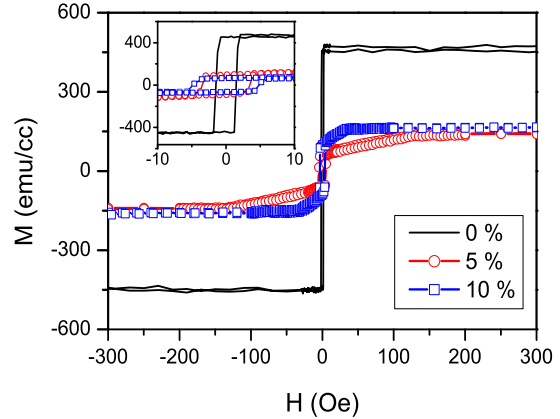}
\caption{\label{fig:Fig11}Magnetization measurements of NiFeN thin
films prepared at different nitrogen partial pressures.}
\end{figure}

The magnetization measurements were also carried out in NiFeN
films as a function nitrogen partial pressure during sputtering.
The M-H loop for the sample prepared with Ar gas only is matching
well with the values obtained for permalloy.\cite{Cullity_MAG} For
the samples prepared with increased R$_{N2}$, the magnetization
decreases rapidly and the values of coercivity increased
(\emph{see} Fig.~\ref{fig:Fig11}).
As will be discussed later, the start structure with NiFe is fcc,
and the volume of interstitial is much larger in case of fcc as
compared with a bcc Fe. This allows more nitrogen atoms to be
incorporated within the unit cell of fcc NiFe, which eventually
results in decrease in magnetization more rapidly as compared with
bcc Fe. The observed magnetization behavior may be understood with
this argument.

\subsection{\label{sec:level36}Polarized neutron reflectivity measurements}

In the present case since the samples were deposited either on
glass or Si substrate and the thickness of the deposited films was
in the range of 100 nm, the diamagnetism of the substrates might
results in erroneous values of absolute saturation magnetization
when measured with DC extraction magnetometer. Also the errors in
determining the size of the measured samples may lead to further
errors in the values of absolute magnetization. With these two
parameters in mind, the magnetization of the FeN samples was also
determined using Polarized neutron reflectivity (PNR). PNR is a
technique which is able to yield the absolute value of magnetic
moment per atom in a magnetic thin film with high
accuracy.\cite{Blundell_PRB92} In contrast to bulk magnetization
magnetometer technique (e.g. DC extraction, VSM or SQUID), no
correction due to magnetic signal from the substrate has to be
applied in PNR. Further, the samples dimensions and mass does not
play a crucial role in determination of magnetic moment. During
the experiment, polarized neutrons with spin parallel or
antiparallel to the direction of magnetization on the sample are
reflected-off the surface of the sample at grazing incidence. The
measurements were performed with an applied field of 400 Oe, which
is sufficient to reach the saturation magnetization in all the
samples. The measurements were carried out in the ToF mode at a
fixed angle of incidence. The ToF-PNR has an advantage as during
the measurement of spin up and spin down reflectivities, only the
polarization of the incoming beam is changed by switching the
direction of the applied field at the polarizing
supermirror.\cite{Gupta_PramJP04} No movement of the sample is
required as often done in $\theta$-$2\theta$ mode. The potential
energy of a neutron in the $i$$^{th}$ region of the sample is
given by:\cite{Blundell_PRB92,Hope_PRB97}
\begin{equation}
{V_{i} = \frac{2\pi\hbar^{2}}{m_{n}}\rho_{i}b_{i} + \mu_{n}\bullet
B_{i}},
\end{equation}
where m$_{n}$, $\rho$$_{i}$, b$_{i}$, $\mu$$_{n}$, B$_{i}$ are
mass of neutron, atomic density, coherent scattering length,
neutron moment and magnetic field. This potential gives rise to
spin dependent reflectivity for the cases when incident
polarization is parallel to the direction of magnetization in the
sample (+) or antiparallel (-).
\begin{figure} [!ht]
\includegraphics[width=80mm,height=100mm]{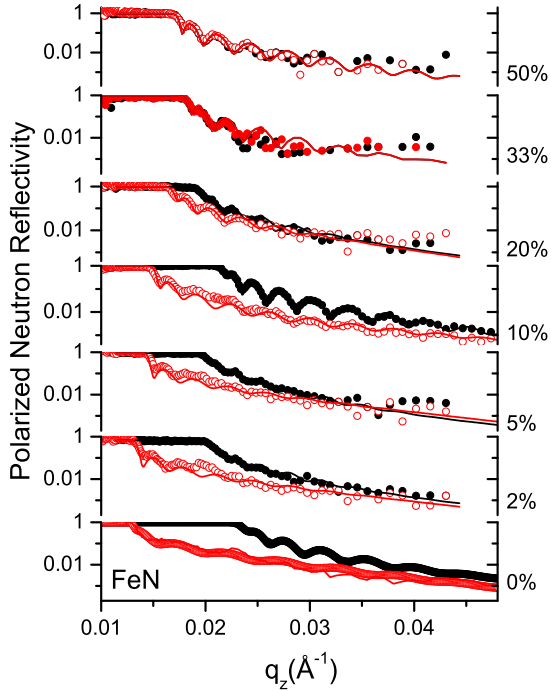}
\caption{\label{fig:Fig12}Polarized neutron reflectivity
measurements of FeN thin films prepared at different nitrogen
partial pressures. Closed circles represents the data
corresponding to spin up reflectivity $R{+}$ and open circles to
spin down reflectivity $R{-}$. The solid line represents fitting
to the experimental data.}
\end{figure}
Fig.~\ref{fig:Fig12} shows PNR data on the samples prepared as
function increased nitrogen partial pressure during sputtering. As
the amount of nitrogen partial pressure is increased, the edge in
$R-$ shows a shift towards higher q$_{z}$ values, and the
separation between $R+$ and $R-$ reflectivities decreases
continuously. Finally for R$_{N2}$$>$20, both $R+$ and $R-$ are
merged together, indicating that the sample has became
nonmagnetic. The magnetic moment in each case was determined by
fitting the experimental data using a computer
program.\cite{SimulReflec}
\begin{figure}
\includegraphics[width=85mm,height=70mm]{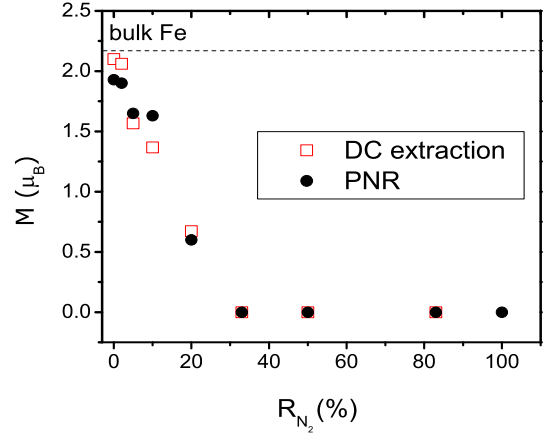}
\caption{\label{fig:Fig13}Magnetic moment as obtained with the
bulk magnetization measurements and polarized neutron
reflectivity.}
\end{figure}
Fig.~\ref{fig:Fig13} shows a plot of obtained values of magnetic
moment with PNR and bulk magnetization measurements. Perusal of
the figures gives a clear indication that the values of magnetic
moment obtained with the two techniques lies in the same range
within the experimental errors. It is interesting to note that at
no point the magnetic moment is higher than that of pure.

\section{\label{sec:level4}Mechanism inducing nanocrystallization or amorphization }

From the results and discussion given in the previous section, it
is clear that both Fe and NiFe form a nanocrystalline or amorphous
structures when sputtered with reactive nitrogen. Only at some
specific nitrogen partial pressures, polycrystalline structures
are obtained. The mechanism inducing nanocrystallization or
amorphization can be understood in terms of incorporation of
nitrogen within the crystal structures of Fe and NiFe and rapid
quenching of adatoms at the substrate. Atoms with low Z e.g. B, C
or N can easily occupy interstitial sites causing an expansion of
the unit cell as well as restricting the long range ordering. It
may be noted that for Fe the structure is bcc, while for NiFe, it
is fcc. The probability of occupying the interstitial sites in the
bcc and fcc crystal system is different because of different close
packing.
\begin{figure} [!ht]
\includegraphics[width=85mm,height=110mm]{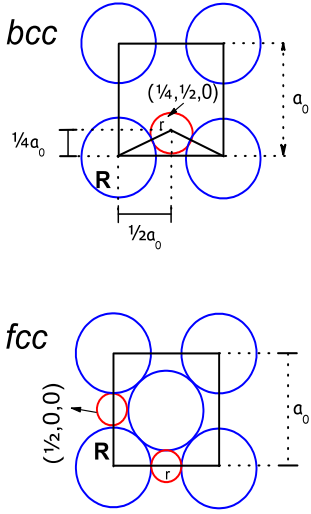}
\caption{\label{fig:Fig14}Distribution of interstitial site in bcc
and fcc structures.}
\end{figure}
In bcc-Fe the atoms may occupy the tetrahedral interstitial sites
(see Fig.~\ref{fig:Fig14}), such as
($\frac{1}{4}$,$\frac{1}{2}$,0), while in a fcc structure,
nitrogen atom may be located either at the octahedral sites at the
edge center of unit cell ($\frac{1}{2}$, 0, 0) and/or at the
center of the unit cell
($\frac{1}{2}$,$\frac{1}{2}$,$\frac{1}{2}$). The probability of
occupation of nitrogen atoms at the interstitial sites in the two
structures can be obtained by calculating the size of the
interstitials for the two cases:

The radius of tetrahedral interstitial site at
($\frac{1}{4}$,$\frac{1}{2}$,0) location in bcc-Fe can be written
as:
\begin{equation}
{r_{int} = [(\frac{1}{2}a_{0})^{2}+(\frac{1}{4}a_{0})^{2}
]^{\frac{1}{2}}-R_{bcc}
 },
\end{equation}
where R$_{bcc}$ is the radius of bcc Fe and $a$$_{0}$ (=0.2866 nm
)is the lattice constant of the bcc Fe atom and R$_{bcc}$ =
$\sqrt{3}$a$_{0}$/4 = 0.1241 nm. From equation (3) the radius of
($\frac{1}{4}$,$\frac{1}{2}$, 0) location of interstitial site
r$_{int}$ for bcc Fe is 0.0361 nm. Similarly we can calculate the
radius of interstitial site for fcc NiFe at the octahedral sites
($\frac{1}{2}$, 0, 0) using the expression:
\begin{equation}
{2r_{int} = a_{0} - 2R_{fcc}},
\end{equation}
Where $a$$_{0}$ = 0.3545nm for NiFe and R$_{fcc}$ is the radius of
fcc NiFe and R$_{fcc}$ = $\sqrt{2}$a$_{0}$/4 = 0.1253 nm. From
equation (4) the radius of ($\frac{1}{2}$, 0, 0) location of
interstitial site $r$$_{int}$ for fcc is 0.0523 nm.

The intestinal site in the bcc Fe is smaller than that in fcc NiFe
alloy. Whereas both are smaller than atomic radius of nitrogen
atom (0.075 nm). Therefore for both Fe and NiFe, nitrogen
occupying the interstitial sites would cause a distortion to the
unit cell and it is expected that this distortion should be more
effective for bcc Fe as compared with fcc NiFe while depositing at
a given nitrogen partial pressure. As a matter fact it is clear
from our XRD results that almost complete amorphization of Fe was
observed at R$_{N2}$ = 10\%, while incase of NiFe, fully amorphous
state was obtained at R$_{N2}$ = 33\%. This result clearly
indicated that nitrogen atoms gradually occupy the interstitial
space within bcc or fcc structure and since available space in a
fcc structure is larger, fcc structure allowed more nitrogen atoms
to be incorporated. Further, when nitrogen partial pressure was
increased beyond the one at which final amorphous structure was
obtained, the structure of both Fe and NiFe was changed basically
to A$_{3}$N (A = Fe or NiFe).

From the energetics of binary iron nitrides\cite{Tessier_SSS00}
(at room temperature), it may be noted that the heat of formation
for $\epsilon$-Fe$_{3}$N$_{x}$ is lower (-40 to -45 kJ mol$^{-1}$)
as compared with neighboring e.g. Fe$_{4}$N (-12 kJ mol$^{-1}$) or
Fe$_{2}$N (-34 kJ mol$^{-1}$) phases. On the other hand nitrogen
richest phase $\gamma$$^{\prime\prime\prime}$-FeN$_{0.91}$ has the
lowest enthalpy of formation (-47 kJ mol$^{-1}$). It is expected
that at intermediate nitrogen pressures when no more interstitial
nitrogen can be incorporated within the unit cell, Fe$_{3}$N or
Fe$_{2}$N phases would be readily formed. In fact when sputtered
with R$_{N2}$ = 33\%, iron nitride structure is a mixture of
Fe$_{3}$N and Fe$_{2}$N phases. Further increase in nitrogen
partial pressure at 50\%, resulted in formation of nitrogen rich
$\gamma$$^{\prime\prime\prime}$-type FeN. Still since the peaks
were broadened, completely crystalline structure was not formed.
At R$_{N2}$ = 83\%, sharp reflections corresponding to
$\gamma$$^{\prime\prime\prime}$-FeN were observed. A further
increase in the nitrogen partial pressure resulted in broadening
of the XRD peaks. This means that well-defined polycrystalline
structure of FeN compounds are only obtained at some specific
nitrogen partial pressures and below and above these specific
partial pressures the long range ordering is restricted due to
incomplete Fe-N bonds or partial breaking of Fe-N bonds due to
excessive nitrogen. Similar behavior was also observed for NiFe,
however amount of incorporation of nitrogen atoms in the two cases
is different.

The energy of the adatoms with parameters used during sputtering
for Fe or NiFe would be around 10 eV,\cite{Turner_JVCTA92} which
corresponds to roughly 10$^{5}$ K. During condensation onto the
substrate which tales place within $\sim$msec, the adatoms are
rapidly quenched, the mobility of the atoms is restricted; it is
expected that either the occupancy of reactive nitrogen at
interstitial sites or a chemical reaction between sputtered atom
and nitrogen takes place in the plasma. Since the substrate were
not heated intentionally, it is very unlikely that any
rearrangement process would take place onto the substrate after
condensation. This argument supports that microstructure of the
deposited film would strongly depend on the plasma which in turn
depend upon the amount of reactive nitrogen used during
sputtering. Since nitrogen atoms gradually occupy interstitial
sites in Fe or NiFe, it is expected that the microstrain of the
structure should increase as the amount of nitrogen partial
pressure is increasing. The microstrain variation in the samples
studied in the present case can be determined from XRD data using
Williamson-Hall plots for Lorentzian peak
shape:\cite{Rosenberg_JPCM2000}
\begin{equation}
{b \cos\theta = A + B\sin\theta}
\end{equation}
Where $A$ = 0.9$\lambda$/$t$, $t$ is grain size and $B$ =
4$\epsilon$($\epsilon$ = microstrain), $b$ is the FWHM in
2$\theta$.

A straight line fit in the plot of $b$ $\cos$$\theta$ and
$\sin$$\theta$ yield the values of constant $A$ and $B$, which in
turn yield the values of grain size as well as microstrain. Such a
plot was obtained all the sample displayed in fig.~\ref{fig:Fig1}
and fig.~\ref{fig:Fig4}.
\begin{figure} [!ht]
\includegraphics[width=85mm,height=75mm]{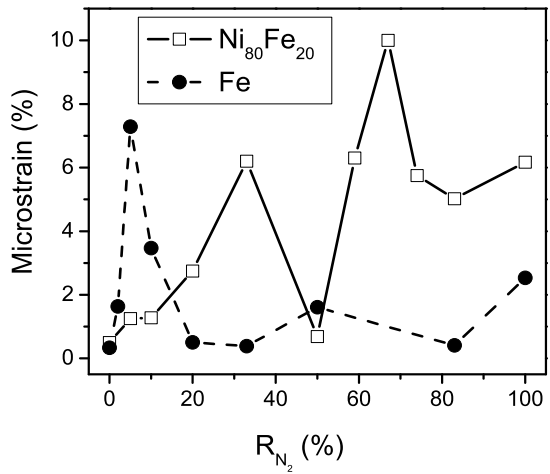}
\caption{\label{fig:Fig15}Microstrain obtained from the XRD data
(see fig.~\ref{fig:Fig1} and fig.~\ref{fig:Fig4})for FeN and NiFeN
samples prepared with different nitrogen partial pressures. The
solid or dotted line are guide to the eye.}
\end{figure}
The obtained values of microstrain for FeN and NiFeN are plotted
in fig.~\ref{fig:Fig15}. It may be noted that the higher order
reflections in the present case are reflected poorly, effective
microstrain is obtained can be obtained using the most intense
peak.\cite{Shaginyan_TSF2002} Further, the microstrain in the
deposited samples can have other origin such as grain size or
dislocation. However a comparison of microstrain for the samples
prepared at different nitrogen partial pressure is expected to
provide additional insight on the mechanism inducing
nanocrystallization or amorphization. It is interesting to see
that for both FeN and NiFeN the microstrain is largest for the
amorphous phases while for polycrystalline phases of nitrides or
pure Fe and NiFe, it is almost zero. An increase in the
microstrain on nanocrystallization and amorphization gives an
indication that the unit cell of Fe or NiFe is distorted due to
incorporation of nitrogen at interstitial sites. Whenever a
polycrystalline structure is formed, the microstrain is at
minimum. Since the polycrystalline structure is formed due to
almost complete covalent bonds between Fe and N or NiFe and N, the
distortion of the unit cell should be at minimum. An incomplete
bonding or excessive nitrogen increased the microstrain in the
film which leads to nanocrystalline or amorphous phase of the
deposited sample. This argument combined with the fact that
adatoms are quenched at the substrate within a very short time,
explains the mechanism inducing nanocrystallization or
amorphization. Even though amorphization or nanocrystallization
increased the microstrain within the structure of the deposited
film, it is interesting to note that surface roughness of the
amorphous film was improved on amorphization, which gives an
indication that strains are developed within the structure.

\section{\label{sec:level5}CONCLUSIONS}
From the present study it can be concluded that reactive nitrogen
sputtering of bcc-Fe and fcc-NiFe (at room temperature) forms a
nanocrystalline or amorphous structure at most of the nitrogen
partial pressures. Only at specific nitrogen partial pressures
polycrystalline nitrides were obtained which are thermodynamically
favored. Above or below this specific pressure the long range
ordering is perturbed by reactive nitrogen. The crystallization in
both amorphous FeN and NiFeN takes place in a single step and at
higher temperature nitrogen poor phases are formed due to nitrogen
out-diffusion. The growth behavior of amorphous FeN showed
improved surface roughness due to amorphization and the growth
exponent $\beta$ was minimum for amorphous phase as compared with
poly or nanocrystalline phases. The magnetic measurements on
ferromagnetic FeN and NiFeN films reveal that in case of FeN, at
low nitrogen content, the alloy forms a soft-magnetic phase while
at higher nitrogen content the average value of magnetization
decreased and coercivity increased. For NiFeN, inclusion of
nitrogen produced phases with reduced values of magnetization. The
magnetic moment of the samples was confirmed with polarized
neutron reflectivity and was in agreement with the values obtained
with DC extraction magnetometery.

\begin{acknowledgments}
We are grateful to Mr. Michael Horisberger for providing help with
thin film deposition and Dr. Yu Lung Chiu for TEM measurements.
One of author (RG), would like to thank Dr. Peter Allenspach for
his support and availing the experimental facilities at Laboratory
for Neutron Scattering, Paul Scherrer Institut.
\end{acknowledgments}

%\newpage %Just because of unusual number of tables stacked at end
\bibliography{FeN_NiFeN_Nano_Amorphous_Gupta}% Produces the bibliography via BibTeX.
\end{document}